\documentclass[preprint,nofootinbib,showpacs,preprintnumbers,amsmath,amssymb]{revtex4}
\usepackage{graphicx}
\newcommand{\bea}{\begin{eqnarray}}
\newcommand{\eea}{\end{eqnarray}}
\newcommand{\be}{\begin{equation}}
\newcommand{\ee}{\end{equation}}

\newcommand{\bi}{\begin{itemize}}
\newcommand{\ei}{\end{itemize}}

\begin{document}
\preprint{}
\title{Reevaluating the Cosmological Origin of Dark Matter}
\author{Scott Watson\footnote{gswatson@syr.edu}}
\affiliation{
Michigan Center for Theoretical Physics, Ann Arbor, MI 
\\
and \\
Department of Physics, Syracuse University, Syracuse, NY 
}
\date{\today}

\begin{abstract}
  The origin of dark matter as a thermal relic offers a compelling way in which
the early universe was initially populated by dark matter. Alternative
explanations typically appear exotic compared to the simplicity of thermal
production. However, recent observations and progress from theory suggest that
it may be necessary to be more critical. This is important because ongoing
searches probing the microscopic properties of dark matter typically rely on
the assumption of dark matter as a single, unique, thermal relic. On general
grounds I will argue that non-thermal production of dark matter seems to be a
robust prediction of physics beyond the standard model. However, if such models are to lead to realistic phenomenology, they must sit in a restrictive theoretical framework. As we will show, as a consequence of such restrictions, viable models will result in concrete and testable predictions. Although many challenges remain, the non-thermal component of such models may offer a new way to test string theories that are formulated to provide realistic particle physics near the electroweak scale.
\end{abstract}

\maketitle
\newpage
\vspace{-1.2cm} 

\section{Cosmological Evidence for Dark Matter \label{section1}}
The first hint for the existence of dark matter came from observations of the nearby Coma cluster of galaxies by Fritz Zwicky in 1933 \cite{zwicky}.  Zwicky found that by assuming the galaxies comprising the cluster were in equilibrium, their velocity distribution implied a cluster mass far exceeding that inferred from the luminous matter contained within the cluster. Today, through a number of complementary and more sophisticated techniques, cluster studies suggest a relative abundance of dark matter $\Omega_{cdm} = 0.2$ to $0.3$, where $\Omega_{cdm}= \rho_{cdm} / \rho_c$ is the fractional amount of dark matter as compared to the critical density for collapse which today is given by $\rho_c=3 H_0^2/(8 \pi G) \approx 10^{-29}$g$\cdot$cm$^{-3}$ with a Hubble parameter of around $70$ km$\cdot$s$^{-1}$$\cdot$Mpc$^{-1}$ and $G=6.67 \times 10^{-11}$ is Newton's gravitational constant.

A more precise measure of dark matter can be obtained from less direct observations, such as the temperature anisotropies of the cosmic microwave background (CMB), and the evolution and formation of the large scale structure (LSS) of the universe.  This is because the evolution of density inhomogeneities that eventually grow to form LSS is quite sensitive to the properties of the primordial bath of particles from which they evolve.  At the time the CMB photons last scattered, by mass the particles were primarily composed of dark matter.  Combining probes of the CMB, structure formation, and distance probes such as supernovae the amount of dark matter is found to be \cite{wmap} 
\be \label{cdm}
\Omega_{cdm}=0.233 \pm 0.013,
\ee
implying that the total energy budget of the universe is comprised of a little less than a quarter dark matter.

In addition to determining the dark matter abundance, these observations, along with the above mentioned galaxy and cluster observations, also tell us that the dark matter must be `cold' -- meaning non-relativistic at the time of structure formation, stable (at least until recently), and `dark' meaning without significant electromagnetic interactions.  The latter, when combined with constraints from Big Bang Nucleosynthesis (BBN), suggests the particles are at most weakly interacting with themselves and with other particles.  Combining all of these cosmological observations, we find that what is expected is a WIMP, that is a Weakly Interacting Massive Particle.

\section{Reevaluating the WIMP Miracle}
\subsection{WIMPs as Thermal Relics  \label{sectionthermal}} 
Big Bang cosmology predicts that as the universe expands it cools\footnote{In this subsection we briefly review the scenario of thermal production of WIMPs.  For a more detailed treatment we refer the reader to \cite{Wells} }.
Thus, if we consider the expansion in reverse, we expect at some point in the early universe that the cosmic temperature would have exceeded the mass of the dark matter particles rendering them relativistic.  At this temperature the particles are relatively light and easy to produce from the primordial plasma so that their creation and annihilation would be near thermal equilibrium.
In equilibrium, the rate at which particles annihilate in a fixed comoving volume $a^3$ is $n_x a^3 \times n_x \langle \sigma_xv \rangle$, where $n_x$ is the number density of dark matter particles of mass $m_x$, $\sigma_x$ is their annihilation cross section, and $\langle \sigma_xv \rangle$ is the thermally averaged cross-section and relative velocity of the particles.  In equilibrium, particle annihilations should be balanced by particle pair-creation and the rate is given by $(n_x^{eq})^2 a^3 \langle \sigma_xv \rangle$, so that the number in a comoving volume is constant.  This is expressed by the Boltzmann equation
\be 
\frac{d{\left(n_x a^3\right)}}{dt}= - a^3 \langle \sigma_xv \rangle  \left[  n_x^2 - \left(n_x^{eq}\right)^2 \right] ,
\ee 
where the first term on the right is dilution due to particle annihilations ($XX \rightarrow \gamma \gamma$), and the second term is the reverse process of particle creation from the thermal bath ($\gamma \gamma \rightarrow XX$). At high temperatures when $T \gg m_x$, we have $n_x^{eq} \sim T^3$ and since $T \sim 1/a$ the last two terms cancel and the particle density simply scales with the expansion. Once the particles become non-relativistic ($m_x \ll T$) then $n_x^{eq} \sim e^{-m_x/T}$ becomes Boltzmann suppressed and particle production becomes negligible, so that the density of particles rapidly drops due to both the expansion and annihilations. Finally, once the number density drops to the point where the cosmic expansion exceeds the annihilation rate per particle $H \gtrsim n_x \langle \sigma_xv \rangle$, the particles `freezeout' and their number per comoving volume 
is
\be \label{freeze}
\frac{n_x}{s} =\left. \frac{H}{s \langle \sigma_xv \rangle} \right\vert_{T=T_f},
\ee
where all parameters appearing in this expression are to be evaluated at the freeze-out temperature $T_f$ and we have introduced the entropy density $s=(2\pi^2/45) g_\ast T^3 \sim 1/a^3$, which gives a more convenient way to define the comoving frame and $g_\ast$ is the number of relativistic degrees of freedom at the time of freeze-out.  The freeze-out temperature can be found from the number density, since one finds that it closely tracks the equilibrium density near freeze-out.  Thus, at freeze-out $n_x \sim n_x^{eq} \sim e^{-m_x/T_f}$, and the mass to temperature ratio at this time is only logarithmically sensitive to changes in the parameters appearing in (\ref{freeze}).  In fact, for thermally produced dark matter associated with weak-scale physics this ratio is typically $m_x/T=25$, with corrections up to at most a factor of two\footnote{Of course, the Boltmann equation can always be solved numerically and one finds good agreement with the analytic argument given above.  See \cite{rocky} for a more thorough discussion.}.

Assuming no significant entropy production following freeze-out, the number of dark matter particles per comoving volume (\ref{freeze}) will be preserved until today resulting in a density of dark matter
\bea \label{criticaldensity}
\Omega_{cdm}(T) \equiv \frac{\rho_{cdm}(T)}{\rho_c} &=& \frac{m_x n_x(T)}{\rho_c}= \frac{m_x}{\rho_c}  \, \left( \frac{n_x(T_f)}{s(T_f)} \right) s(T) \nonumber \\
&=& \frac{m_x}{\rho_c} \left( \frac{H}{s \langle \sigma_xv \rangle}  \right)_{T=T_f} s(T). 
\eea
Making the additional assumption that the universe is entirely radiation dominated at freeze-out so that $H\sim T^2$ and using $s \sim T^3$ we find that the critical density in dark matter evaluated today is
\bea \label{cdm2}
\Omega_{cdm}(T_0)&=&\frac{45}{2 \pi \sqrt{10}}  \left( \frac{s_0}{\rho_c m_p}\right)  \left(  \frac{m_x}{g_\ast^{1/2} \langle \sigma_xv \rangle T_f  } \right), \nonumber \\
&\approx&0.23  \times \left( \frac{10^{-26} \,  \mbox{cm}^{3} \cdot \mbox{s}^{-1} }{\langle \sigma_xv \rangle} \right), 
\eea
where $g_\ast=106.75$ is the number of relativistic degrees of freedom around the typical temperature of dark matter freezeout (see \cite{Wells} for a more detailed discussion), and the entropy density today is $s_0=2970$ cm$^{-3}$,

This result is interesting for several reasons. 
First, we note that the abundance only depends on the self annihilation cross section of the dark matter particles, and we saw that any changes in the theory enter as logarithmic corrections -- i.e. this scenario is robust.  Thus, measurements of the thermal relic density won't lead to any deeper understanding of physics beyond the standard model or the evolution of the universe prior to freezeout.  This will be an important difference from the non-thermal case that we will discuss below.  Another interesting fact about the result above is that if we compare this result for the abundance of dark matter produced thermally with the precision cosmological measurement given in (\ref{cdm}), we find that $\langle \sigma_xv \rangle \approx 10^{-26} $cm$^{3} \cdot $s$^{-1}$ (or $\sigma_x\approx 1$ picobarn) and thus we are lead to expect a new particle with weak scale interactions. Of course, we already expect new physics to appear near the electroweak scale to properly account for a light Higgs.  Such theories for an extension of the standard model postulate new symmetries above the electroweak scale, and at low energy their breaking results in a lightest stable particle associated with the new physics.  One example is provided by the supersymmetric (SUSY) neutralino, which after the spontaneous breaking of SUSY remains stable under a residual discrete symmetry, i.e. R-parity.
That the weak scale cross section naturally emerges when comparing the cosmological observations with the thermal prediction (\ref{cdm2}), and the fact that this was independently expected from theoretical considerations related to the Higgs has lead some to refer to this coincidence as the `WIMP Miracle'. To summarize, 
\\

Assuming:
{\bi
\item The WIMPs were at some point relativistic and reached chemical equilibrium.
\item At the time of freeze-out, the universe was radiation dominated (all other contributions to the energy density were negligible).
\item Following freeze-out there was no significant entropy production.
\item There were no other late-time sources of dark matter particles (e.g. decays from other particles).
\item There is only one species of dark matter particle and any other new particles are unstable or have significantly larger mass.
\ei}
we find that
{ 
\bi
\item The relic density does not depend on the expansion history, only on the temperature at freeze-out.
\item The relic density does not depend on any high scale physics, only on the low-energy cross-section.
\item The answer is very robust to changes in the cross-section and mass of the particles.
\item When combined with cosmological observations -- we expect new physics at the electroweak scale.
\ei
}

Although all the assumptions listed above are well motivated -- and the resulting model is quite simple and compelling --  it is important to proceed with caution when attempting to promote any candidate signature coming from particle experiments to a claim that one has gained a complete understanding of cosmological dark matter.  In addition to the challenge of reconstructing the properties of dark matter from signatures at colliders, direct, and indirect detection, there are also a number of challenges associated with the reconstruction of the relic density of dark matter itself.  These include that the relic density could be comprised of more than one kind of particle, that the expansion history prior to BBN could be more complicated than expected, or that the late decay of particles could alter the abundance of dark matter particles.
These are just a couple possibilities that could stymy the extrapolation of a confirmed particle detection to an accurate picture of cosmological dark matter. 

\subsection{Other Dark Matter}
One key assumption underlying the connection between the thermal relic abundance (\ref{cdm2}) and LHC, is that the WIMP is a unique dark matter candidate and that its mass is far below the next to lightest particle associated with new physics.  
As an example of the latter, in supersymmetric theories it is common that the next to lightest SUSY particle (NLSP) can be nearly degenerate in mass with the LSP.
If this is the case, not only could the NLSP be mistaken as a stable WIMP (LSP) in the LHC detector -- as the lifetime of a particle in the detector is only $10^{-8}$ s, or the NLSP and its decay products might both be neutral -- but cosmologically, coannihilations \cite{coann} between the NLSP and LSP will significantly reduce the thermal relic density estimated in (\ref{cdm2}).

Another important possibility is that there is more than one type of dark matter.  
Thus, the total dark matter abundance should always be thought of as
\be
\Omega_{cdm}^{total} = \sum_i \Omega_{cdm}^{(i)},
\ee
where the sum is over all contributions to the dark matter energy budget.
In fact, because we now know that neutrinos have mass, we also know that they must make up some part of the dark matter.  
However, we also know that neutrinos are relativistic at the onset of structure formation, i.e. they are  `warm' dark matter, requiring that they must represent a small fraction of the total dark matter.  In fact, combining the recent WMAP5 data with other cosmological observations a bound of  $\Omega_\nu h^2 \lesssim 0.006$ was obtained in \cite{wmap}. Of course, in addition to neutrinos there are a number of other possible contributions to the cosmological dark matter, including axions.  The QCD axion provides an elegant solution to the strong CP problem, and although tightly constrained, still remains a viable dark matter candidate for some regions of its parameter space (see e.g. \cite{arxiv:0807.1726}).
It is also expected that additional axions will generically arise at low energies from effective theories with ultraviolet completions in string theory (see \cite{Svrcek:2006yi} and references within). 

\subsection{Modified Expansion History at Freeze-out}
For the calculation of the thermal relic density one assumption was that the universe was radiation dominated at the time of freeze-out, so that $H \sim T^2$
allowing for the simplification in going from (\ref{criticaldensity}) to  (\ref{cdm2}).  This assumption agrees with the observational predictions of BBN occurring a few minutes after the Big Bang.  However, there is no cosmological evidence for this assumption prior to the time of BBN.  

There are both theoretical and observational indications that this assumption may be too naive.
Indeed, given the rich particle phenomenology that occurs at energies above the scale of BBN (energies around a MeV), we might expect this to complicate the simple picture of a purely radiation dominated universe.
Moreover, relics from early universe phase transitions, such as scalar condensates or rolling inflatons that didn't completely decay, would also be expected to alter the expansion history.

In fact, theories beyond the standard model generically predict the existence of scalar fields. Many of these fields have little or no potential -- so called moduli, so they are often light.  Examples include the sizes and shapes of extra dimensions, or flat directions in the complicated SUSY field space of the scalar partners to standard model fermions.  
In the early universe these moduli will generically be displaced from their low energy minima during phase transitions, such as inflation \cite{Dine:1995kz}.  Energy can then become stored in the form of coherent oscillations forming a scalar condensate.   The cosmological scaling of the condensate depends on which term in the potential is dominant.  For a potential with a dominant term $V \sim \phi^\gamma$ one finds that the pressure depends on the energy density as
\be \label{pressure}
p=\left( \frac{2 \gamma}{2+\gamma} - 1 \right) \rho,
\ee
where $\rho$ scales as
\be
\rho = \rho_0 a^{-6 \gamma / (2+\gamma)}.
\ee

Two examples are a massive scalar with negligible interactions for which $\gamma=2$ and the condensate scales as pressure-less matter $p=0$, whereas if physics at the high scale is dominant -- in the form of non-renormalizable operators -- then $\gamma > 4$ and the condensate evolves as a stiff fluid $p \approx \rho$ for large $\gamma$.
Whatever the behavior of the condensate, if it contributes appreciably to the total energy density prior to freeze-out the abundance (\ref{cdm2}) will be altered.  This is because the presence of addition matter will increase the cosmic expansion rate allowing less time for particle annihilations prior to freeze-out\footnote{Here we have assumed that radiation contributes substantially to the total energy density or that whatever the primary source of energy density it scales at least as fast as radiation.  However, if instead the universe were completely dominated by a massive, non-interacting scalar condensate then this would actually decrease the amount of dark matter.  In either situation, the point is that the standard thermal relic density (\ref{cdm2}) will not give the correct result.}.
The expansion rate at the time of freeze-out is then given by
\be
H_f = H_{rdu} \left(1 + \frac{\rho_\phi}{\rho_r}  \right)^{1/2},
\ee
where $H_{rdu}$ is the expansion rate in a radiation dominated universe and $\rho_\phi$ and $\rho_r$ are the energy density of the scalar condensate and radiation, respectively.
Using that at freeze-out $\rho_r = (\pi^2/30) g(T_f) T_f^4$, $\rho_\phi= \rho_{osc} \left(  T_f /T_{osc} \right)^p$ where $p \equiv 6 \gamma / (2+ \gamma)$, and $\rho_{osc}$ is the energy initially in the condensate which began coherent oscillations at temperature $T_{osc}$ we find that the new dark matter abundance is 
\be
\Omega_{cdm} \rightarrow \Omega_{cdm}  \sqrt{1+r_0 T_f^{2(\gamma-4)/(2+\gamma)}} ,
\ee
where we have used $a \sim 1/T$ for an adiabatic expansion, and the constant 
$$
r_0 \equiv \frac{30}{\pi^2} \left(\frac{\rho_\varphi(T_{osc}) }{ g(T_f) T_{osc}^{6\gamma/(2+\gamma)} }\right),
$$ 
where $g(T_f)$ is the number of relativistic degrees of freedom at freeze-out.
In practice, typically one finds that for moduli in the early universe $r_0 \gg 1$ \cite{Dine:1995kz,Acharya:2008bk}.
We see that especially for high energy effects in the potential this can have a significant effect on the resulting relic density.
As a simple example, if we consider a massive scalar with negligible interactions ($\gamma=2$) displaced after a period of inflation we expect $\rho_\varphi(T_{osc}) \simeq m_\varphi^2 m_p^2$ so that $r_0 \simeq (m_\varphi m_p)^2 / (g(T_f) T_{osc}^3) \gg 1$ leading to a large enhancement of the relic density.
One can also show that there is a significant effect for scalars which are dominated by their kinetic terms (e.g. kination models \cite{Salati:2002md,Catena:2004ba,Chung:2007vz,Chung:2007cn} ), which 
behave like the stiff fluid models discussed above (i.e. $p=\rho$).   In fact, this modification to the expansion history was considered in \cite{Grin:2007yg}, where it was shown that this would loosen constraints on axionic dark matter.
In these examples, the relic density is found to be enhanced compared to that of a purely radiation dominated universe.
Of course scalar condensates are not the only additional sources of energy one might expect in the early universe and it is important to note that any additional, significant component will alter the standard thermal abundance of the cosmological dark matter in a way similar to that discussed for scalars above.   

\subsection{Late Production of Dark Matter and Entropy}
Two more crucial assumptions that went into the dark matter abundance (\ref{cdm2}) were that there were no other sources of dark matter and/or entropy production following freeze-out. An example of how this can fail is if there is a late period of thermal inflation \cite{Lyth:1995ka}, which has been argued to be quite natural and necessary for resolving issues with some models coming from string compactifications (see e.g. \cite{Conlon:2007gk}). Another example is provided by the condensate formation we discussed above. 
That is, because the moduli have very weak couplings -- typically of gravitational strength -- the condensate will decay late producing additional particles and entropy.  This decay must occur before BBN, which requires the modulus to have a mass larger than around $10$ TeV in order to avoid the so-called cosmological moduli problem
 \cite{Coughlan:1983ci,Ellis:1986zt,deCarlos:1993jw,Banks:1993en}.  If the condensate contributes appreciably to the total energy density at the time it decays it will not only produce relativistic particles -- and significant entropy -- but could also give rise to additional dark matter particles.  The former will act to reduce the thermal relic density of dark matter particles $\Omega_{cdm} \rightarrow \Omega_{cdm} \left(  T_r / T_f \right)^3$, where $T_r$ is the temperature after the decay and $T_f$ is the freeze-out temperature of the dark matter particles. The factor by which the abundance is diluted can be understood from the scaling of the volume $a^3$ and we have $T\sim 1/a$. As an example, for a $10$ TeV scalar the decay to relativistic particles will `reheat' the universe to a temperature of around an MeV, whereas a $100$ GeV WIMP freezes out at a temperature near a GeV.  Thus, the scalar decay will dilute the preexisting relic density in dark matter by a factor of about $(T_r/T_f)^3 \simeq 10^9$.  

As we have mentioned, in addition to the scalar decaying to relativistic particles it could also decay to WIMPs below their freeze-out temperature.  In this case there are two possible results for the relic density, depending on the resulting density of the WIMPs that are produced \cite{Moroi:1999zb}.   If the number density of WIMPs exceeds the critical value
\be \label{ntprod}
n^c_x =\left. \frac{H}{ \langle \sigma_xv \rangle} \right\vert_{T=T_r},
\ee
then the WIMPs will quickly annihilate down to this value, which acts as an attractor.  It is important to note that the fixed point value is evaluated at the time of reheating, in contrast to the freeze-out result (\ref{freeze}).  The other possibility is that the WIMPs produced in the decay do not exceed the fixed point value.  In this case their density is just given by $n_x \sim B_x n_\varphi $, where $B_x$ is the branching ratio for scalar decay to WIMPs and $n_\varphi $ is the number density of the scalar condensate.
We see that in both these cases the thermal relic density (\ref{cdm2}) would give the wrong answer for the true abundance of dark matter, unless the entropy diluted thermal density of dark matter still manages to exceed the amount coming from the scalar decay.  We see that in the case of fixed point production, comparing (\ref{ntprod}) with the calculation for thermal production (\ref{cdm2}) results in a parametric enhancement proportional to the ratio of the freezeout to the reheat temperature, i.e.
\be
\Omega_{cdm} \rightarrow \Omega_{cdm} \left( \frac{T_f}{T_r} \right),
\ee
which for the example of a $10$ TeV scalar results in an overall enhancement of about three orders of magnitude.  

Fixing the relic density by the cosmological data (\ref{cdm}) implies that  particles need a larger cross section in order to get the right amount of dark matter.  For example, in the case of neutralino dark matter, Winos and Higgsinos  annihilate well and have been seen as giving too little dark matter given a thermal history.  However, in the theoretically constructed models of \cite{Moroi:1999zb,Acharya:2008bk} it is found that Winos, Higgsinos, or some mixture can yield the right amount of dark matter due to non-thermal production, which results naturally by requiring consistency of the theory.

It is important to note that even though the naive thermal freezeout calculation no longer determines the relic density, in the fixed point case the answer is still given in terms of the weak scale cross section, and gives a result of the correct order of magnitude for WIMPs with masses of order $100$ GeV. The dark matter scale and the electroweak symmetry breaking scale still remain related, and the ``WIMP Miracle'' survives.  

\centerline{---------------------------}

In this section, we have seen three possible ways in which the prediction for the amount of cosmological dark matter -- and the constraints on microphysics that would result -- can be altered.  Although the case for more than one type of dark matter, or a more complicated expansion history prior to BBN might seem plausible, the case for a scalar with a mass light enough to decay after dark matter freezeout, but heavy enough to avoid BBN constraints naively would seem quite contrived.  In the next section we will argue that this is not the case, and that hints from model building in a way that is consistent with UV physics might predict that non-thermal production of dark matter is the rule rather than an exotic exception.

\section{Non-thermal Production of WIMPs}  
Non-thermal production of dark matter is not a new idea \cite{nonthermal,Moroi:1999zb,Acharya:2008bk,Giudice:2000ex,Jeannerot:1999yn,ps}.  However, recent results and future expectations from both theory and experiment suggest that such an origin for dark matter might need to be seriously reconsidered.  On the observational side, cosmic ray experiments such as PAMELA and FERMI have reported an excess in both cosmic ray positrons and gamma rays above anticipated astrophysical backgrounds.  Although a dark matter explanation seems somewhat unlikely, if the dark matter had a larger cross-section -- as made possible by non-thermal production -- then candidates like the Wino neutralino may be capable of addressing the excesses through the self annihilations of dark matter \cite{Kane:2009if,Grajek:2008pg,arxiv:0807.1508,arxiv:0807.1634}.  Conversely, current and future data from experiments like PAMELA and FERMI can be used to put important constraints on the dark matter cross section and therefore the non-thermal production process \cite{Kane:2009if,Grajek:2008pg,arxiv:0807.1508,arxiv:0807.1634}.  By itself these results are certainly not a compelling argument for non-thermal production, but another motivation could be provided in the very near future by the Large Hadron Collider (LHC) or other future colliders.  That is, if dark matter was non-thermally produced resulting in a larger self annihilation cross-section, then cross sections for dark matter particles deduced from LHC -- when used to calculate the thermal relic density -- would result in an unacceptably low cosmological abundance and would be in surprising disagreement with e.g. the WMAP data \cite{Kane:2008gb}.  Of course, the explanation could also lie elsewhere, e.g. as a consequence of more than one dark matter particle. Thus, we are lead to the possibility of a `dark matter inverse problem' \cite{Kane:2008gb} -- stressing the importance of combining collider, astrophysical (direct/indirect detection), and cosmological probes in order to obtain a complete understanding of both the microscopic and cosmological nature of dark matter.

\subsection{Considerations from Fundamental Theory}
Given the possible observational consequences of non-thermal production, it is important to ask if such a scenario makes sense from a fundamental viewpoint, or whether such models represent exotic physics.
We saw in the last section that interesting (meaning leading to a situation different from thermal production) and viable cases of non-thermal production rely on {\bf three crucial assumptions} in order for the WIMP Miracle to survive: 
\bi
\item{A scalar condensate composed of particles with masses of about $10-100$ TeV}
\item{Gravitational coupling to all matter}
\item{A new symmetry that when broken leads to a stable dark matter candidate}
\ei
All of these requirements are a natural consequence of physics beyond the standard model.  However, the very particular choice of an approximately $10$ TeV scale mass for the decaying scalar -- though mandatory -- seems quite artificial.  That is, if the scalar is lighter than about $10$ TeV then it threatens the successes of BBN, whereas if it is much heavier it would decay before dark matter freezeout and we would have the usual thermal dark matter scenario.  It is this apparent tuning of the scalar mass that makes the scenario of non-thermal production much less aesthetically appealing than the thermal case which appears quite robust.  Indeed, from a phenomenological point of view it is hard to motivate such a scalar mass except in special cases (see e.g. \cite{Moroi:1999zb}), however the scenario does have the advantage of being testable in current and near term experiments as discussed above.

This picture drastically changes if one considers constructing phenomenological models which are theoretically consistent in the presence of gravity and at high energies, i.e. for models which have a UV completion in quantum gravity.  
At first, decoupling of scales would seem to suggest that high energy physics -- far beyond the scale of electroweak symmetry breaking -- should be irrelevant for the low energy physics of dark matter and the standard model. 
However, string theories, while providing a consistent UV completion, also provide a very rigid set of constraints that must be applied to low energy effective field theories (EFTs) that would otherwise seem perfectly consistent at low energies and in the absence of gravity \cite{vafanima}.
In this way one can hope to highly constrain the number of possible phenomenological models, using added constraints resulting from demanding consistency conditions, such as the absence of anomalies in the presence of gravity \cite{vafanima}.
String theory provides a framework to build such models, however, whether one uses string theory or some other consistent UV completion 
a successful top-down approach must {\bf at least} provide: 
\bi 
\item{A four dimensional effective theory containing a perturbative limit in which we recover the standard model and Einstein gravity.}
\item{An explanation for the hierarchy between the Planck scale and the scale of electroweak symmetry breaking.}
\item{Additional symmetries must be spontaneously broken -- as to not reintroduce the hierarchy problem.}
\item{The vacuum should contain a small and positive cosmological constant (or equivalent) today.}
\ei
Although at this time no single theory has been shown to accomplish all of these goals in a convincing and natural way, it is interesting that in string theories all these problems can be related to the problem of stabilizing light scalars -- moduli.  
These moduli parameterize the structure of the vacuum of the theory.  They describe the size and shape of extra dimensions, as well as the location and orientation of any strings and/or branes that are present.  In addition, at the phenomenological level, scalars will also appear as the superpartners to the standard model fermions and many of these scalars lead to flat directions in the potential, i.e. directions in field space where no forces act.  

Given the expectation of a large number of scalars with little or no potential, it has been an important program in string model building to find ways in which these scalars may have been stabilized, or at least ways in which the formation of scalar condensates might have been prevented\footnote{See e.g. \cite{McAllister:2007bg} for a guide to the literature.}.  This is crucial to avoid the cosmological moduli problem discussed above.  

An essential step in the program to stabilize the vacuum was the inclusion of additional string theoretic ingredients, which were naturally expected to appear in the theory, but had been neglected initially for computational simplicity. 
It was later found that the inclusion of branes, strings, and generalizations of Maxwell fields (fluxes) lead to stabilizing effects that ultimately lead to string scale masses for many of the scalars.  It then follows that these extremely heavy particles would quickly decay in the early universe to lighter particles, and we have an effective decoupling of string scale physics as one would naively expect.

The low energy, four dimensional scalar potential is then given by
\be \label{sp}
V=e^{K/m_p^2}  \left( \sum_\alpha \left| D_\alpha W  \right\vert^2 - 3 \frac{|W|^2}{m_p^2} \right)
\ee
where the sum runs over all fields present in the low energy theory, $W$ is the superpotential, $K$ is the Kahler potential, and the condition for SUSY is that $D_\alpha W \equiv \partial_\alpha W + W \partial_\alpha K$=0.  The stabilization of the moduli at high energy leads to a constant term in the low energy superpotential $W=W_0$.  For a generic choice of flux, SUSY will be broken explicitly and at the string scale.  However, if we choose flux that preserves SUSY, then (\ref{sp}) with $D_\alpha W=0$ implies a deep, negative potential leading to an AdS or negative cosmological constant vacuum.
In order to break SUSY and lift the potential we must add an energy contribution to the potential that is parametrically of the form \cite{deCarlos:1993jw}
\be 
\Delta V(\Phi) = m_{3/2}^2 m_p^2 f\left( \frac{\Phi}{m_p}\right),
\ee
where $m_{3/2}$ is the gravitino mass, related to the scale of SUSY breaking by $\Lambda_{SUSY}^2= m_{3/2} m_p$, and $\Phi$ is the field leading to the symmetry breaking.
The inclusion of physics that would lead to a term like that above is restricted if we hope to achieve a realistic and successful theory.  It must lead to {\em spontaneous} SUSY breaking and a gravitino mass of $m_{3/2} \approx$ TeV, if it is to preserve the success of SUSY in explaining the scale of electroweak symmetry breaking.  This is important since $m_{3/2}$ sets the mass of the superpartners and these can not be far above the electroweak scale.  It must also cancel the contribution on the right side of (\ref{sp}) arranging for a small positive cosmological constant.

It might be difficult to understand how a string based model could ever accomplish this given the discrepancy of scales. However, in addition to the stabilized scalars, the presence of additional symmetries in the theory generically leads to the situation that at least one (if not many) of the scalars are not stabilized at the perturbative level\footnote{In some cases this is tied to the requirement that the resulting low energy theory must be perturbative (small coupling), and since the expectation values of many of these scalars determine the low energy couplings, this forces their stabilization away from the string scale. (see e.g. \cite{dineus}).}.  For these scalars it was shown that non-perturbative effects, such as the condensation of fermions (gauginos in a strongly coupled hidden sector) \cite{Nilles:1982ik}, or the presence of additional branes \cite{Kachru:2003aw} or additional hidden sector matter fields \cite{Lebedev:2006qq} can be used to stabilize the remaining scalars, providing them with a mass.  This leads to an additional contribution in the superpotential and we have
\be
W=W_0+m_p^3 e^{-X},
\ee
where for simplicity we consider the case of a single scalar $X$ and we take the string scale to lie near the Planck scale -- these assumptions however are not crucial to the arguments to follow.
The Kahler potential is then of the form
\be
K=-n m_p^2 \log \left(  X+\bar{X} \right).
\ee
The SUSY minimum corresponds to
\be
D_XW=0 \rightarrow \langle X \rangle = \log \left( \frac{m_p}{n m_{3/2}} \right)
\ee
and using this in (\ref{sp}) we again find the AdS minimum
\be
V_{\mbox{AdS}}=-3 m_{3/2}^2 m_p^2,
\ee
which although SUSY preserving, we choose to write in terms of the gravitino mass in anticipation of SUSY breaking. 
The authors of \cite{Kachru:2003aw} then argued that one could break SUSY and lift the vacuum to contain a small cosmological constant by the addition of another brane leading to a contribution to the potential
\be
\Delta V \sim m_{3/2}^2 m_p^2.
\ee
It is important to mention that such an addition must meet rigid constraints coming from the high energy theory that are required for the consistency of the theory (tadpole/anomaly cancelation).
Given the full potential we can canonically normalize the scalar field 
\be
\delta X \rightarrow \delta X_c = \frac{\sqrt{n}}{\langle Re X \rangle} \delta X
\ee
and we find that its mass is then given by
\be
m_X = \frac{1}{\sqrt{n}} \log \left( \frac{m_p}{n m_{3/2}}  \right) m_{3/2}.
\ee
This scaling and its relation to phenomenology was first stressed in \cite{LoaizaBrito:2005fa}.
We see that in order to preserve the hierarchy one would need $m_{3/2} \approx TeV$ and so the scalar mass would naturally lie near the TeV scale. Of course this result just demonstrates that if a scalar of string origin is protected under a symmetry until SUSY breaking occurs its mass should be on the order of the gravitino mass, which must be near a TeV for naturalness. This is precisely the result needed for the non-thermal production of dark matter to be natural, suggesting a new `non-thermal' WIMP miracle\cite{ps}.

Of course the scenario mentioned above is very far from realistic.  First, the model of \cite{Kachru:2003aw} would seem to explicitly break SUSY by the addition of the brane, where a realistic model should spontaneously break the symmetry.  However, this point is moot, because the model contains two tunings -- one for the cosmological constant and one for the gravitino mass.  The latter implies that the phenomenological successes of SUSY are lost.   
To see this, consider the gravitino mass in the theory which is given parametrically by
\be
m_{3/2}=\frac{|W_0|}{m_p^2 V_6},
\ee
where $V_6$ is the overall volume of the extra dimensions.  In the models of \cite{Kachru:2003aw}, one then tunes the values of the flux to yield a small value for the superpotential ($W_0 \ll 1$) and thus the scale of SUSY breaking.  Another class of models, so-called Large Volume models \cite{Conlon:2005ki}, take the natural value $W_0 \approx 1$, but then tune the volume\footnote{The authors of \cite{Conlon:2005ki} argue that this is not a tuning but the natural location when considering higher order corrections to the theory.  This remains to be seen however, since it is difficult to systematically calculate all corrections to the theory.} $V_6 \approx 10^{14} \gg 1$ so as to obtain the correct scale of SUSY breaking.  Another possibility arises from considering M-theory compactifications \cite{Acharya:2006ia} where it is argued that all moduli are stabilized by non-perturbative physics, so that there is no constant contribution to the superpotential ($W_0=0$).  The geometry of these compactifications is quite complicated and offers a substantial challenge, however if the expectation holds this would realize SUSY breaking dynamically \cite{Witten:1981nf} and preserve the hierarchy.  These models also predict the existence of a TeV scale scalar mass, which has been shown to give rise to a non-thermal scenario \cite{Acharya:2008bk}. 

It must be stressed that all of these models contain shortcomings and substantial challenges to address, but with our current understanding of moduli stabilization and SUSY breaking, it would seem that a scalar with TeV mass is an inevitable prediction of the theory.  Of course, this was also the original motivation for the cosmological moduli problem, which was argued to be very robust given the arguments presented above.

\section{Conclusions}
In this review we have seen that dark matter as a thermal relic remains a simplistic and convincing explanation for the cosmological origin of dark matter.  We have also seen that there are a number of possible ways in which this paradigm could turn out to be too naive.  Recent observations from dark matter experiment suggest that this might be the case, but taken alone are not especially compelling.  However, when combined with theoretical expectations, the possibility of non-thermal dark matter seems worthy of serious consideration. This is especially true since it would make concrete predictions for LHC -- if we calculate the thermal relic density from the self annihilation cross section of dark matter deduced from LHC alone we would get disagreement with cosmological observations.  We also saw that the existence of light scalars associated with physics beyond the standard model naturally predicts the existence of a scalar with TeV scale mass -- the essential ingredient for non-thermal production.  This is intimately tied to the cosmological moduli problem, and progress in string theories in addressing this problem suggests that a non-thermal origin of dark matter may be inevitable.  However, model building is in an early stage and there are many challenges that remain in building more realistic models that are compatible with both the standard model and at higher energy with quantum gravity.  Regardless of the outcome of the theoretical effort, if we are to achieve a complete understanding of dark matter (both microscopic and macroscopic) this will require combining collider, astrophysics (direct and indirect), and cosmological observations with theoretical approaches.

\section*{Acknowledgments}
I would like to thank Gordy Kane for discussions, collaboration, and initially suggesting to me to explore many of the ideas presented in this review.
I would also like to thank Bobby Acharya, Konstantin Bobkov, Sera Cremonini, Dan Feldman, Phill Grajek, Piyush Kumar, Ran Lu, Dan Phalen, Aaron Pierce, and Jing Shao for discussions and collaboration.   The research of S.W. is supported in part by the Michigan Society of Fellows.  S.W. would also like to thank Cambridge University -- DAMTP and the Mitchell Institute at Texas A\&M for hospitality and financial assistance.

\end{document}